# CDF Run II Status and prospects

C. Pagliarone

*INFN-Pisa, V. Livornese 1291, 56010 Pisa, Italy*

*e-mail: pagliarone@fnal.gov*

(On the behalf of the CDF Collaboration)

Abstract

Run II at the Tevatron Collider started at the beginning of March 2001. With extensive upgrades on both detectors and electronics the CDF II began to collect data. This paper reviews early Run II physics results obtained by analyzing data collected before the middle of june 2002. At the present the understanding of the detector performances is rather high so many analysis are already underway.

## 1. Run II Tevatron upgrades

The Fermilab Tevatron Collider has undergone, in the past few years, a whole series of upgrades to increase the instantaneous luminosity and to improve the collider bunch structure. During the first phase of the Run II (Run IIA) the machine is expected to deliver to each of the two collider experiments: CDF and D$\varnothing$, a goal luminosity of up to $5 \div 8 \times 10^{31}$ cm$^2 \cdot$s$^{-1}$ with a center of mass energy of 1.96 *TeV* that is a bit larger than the 1.8 *TeV* of the Run I.

During the present Run IIA the Tevatron is operating much like in the previous Run IB with a higher integrated luminosity mostly coming from an increase in the number of bunches and slightly higher proton and antiproton bunch intensities. The planned integrated luminosity expected by the end of Run IIA is fixed to $\approx 2$ $fb^{-1}$.

The bunch structure of the Tevatron Collider have been changed. Indeed, we passed from the 6×6 proton-antiproton bunches of the Run I to the present 36×36. The replacement of the Main Ring with the Main Injector as the injection source for the Tevatron collider, leaded to an increased number of protons per store and at the same time eliminated a source of background for the detectors. Several upgrades also increased the number of antiprotons per store. New Main Injector creates antiproton beam with higher intensity and energy than in the Run I. In addition, the plan is to recycle 'unused' antiprotons at the end of a collider store rather than dump them. When the Recycler will be fully operational we expect to reach instantaneous luminosity up to $2 \times 10^{32}$ cm$^2 \cdot$s$^{-1}$.

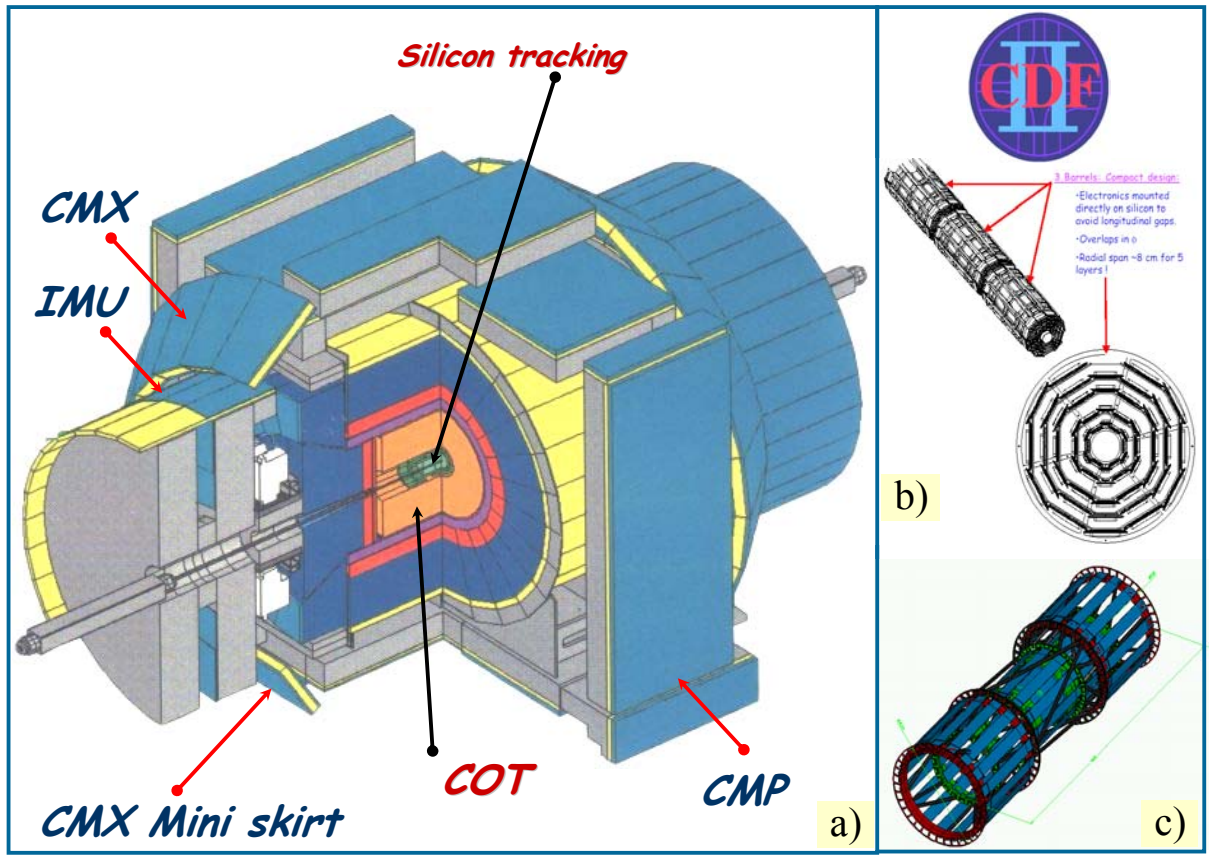

**FIG. 1. a)** *3-Dimensional view of the CDF II detector configuration; the cutaway view of half section of the inner portion of the CDF II detector shows the inner tracking region surrounded by solenoid, endcap calorimeters and, in the most external part, by the muon system (CMP, CMX and IMU);* **b)** *view of the SVX detector;* **c)** *view of ISL detector.*

## 2. The CDF II Detector Improvements

Both Tevatron Collider Detectors have been improved in order to operate with the new machine performances that means mainly with an increased instantaneous luminosity as well as the critical bunch spacing structure. In addition, there have been several upgrades to increase the sensitivity of the detector to specific physics tasks such as heavy flavor physics, Higgs boson searches and many others. A detailed description of the CDF detector upgrades may be found in the following documents [1], [2]. Figure 1.a shows a 3D cutaway view of the final configuration of the CDF II experiment. The central tracking volume of the CDF II experiment has been replaced entirely with new detectors, the central calorimeters has not been changed, the muon system has been mainly increased in coverage. These upgrades can be summarized as follow:

1. **Silicon tracking system** done of 3 different tracking detector subsystems:
    *Layer00* is a layer of silicon detectors installed directly on the beam pipe to increase impact parameter resolution.

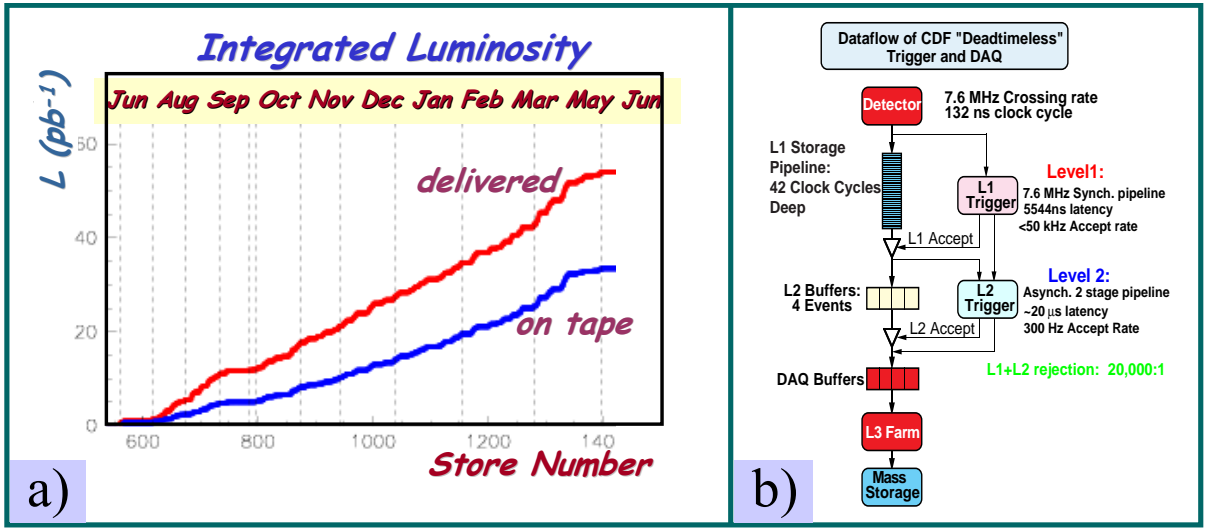

**FIG 2. a)** *Tevatron Run IIA integrated luminosity history through June 2002;* **b)** *Diagram of the CDF II trigger architecture.*

***Silicon Vertex Detector*** (***SVX II***): in order to meet new physics goals, a central vertexing portion of the detector called SVX II was designed (see fig. 1.b). It consists of double-sided silicon sensors with a combination of both 90-degree and small-angle stereo layers. The SVX II is nearly twice as long as the original SVX and SVX' (*96 cm* instead of *51 cm*), which were constrained to fit within a previous gas-based track detector (VTX), used to locate the position of interactions along the beam line. SVX II has 5 layers instead of 4 of the two previous silicon vertex detectors and it is able to give 3-dimensional information on the tracks.

***Intermediate Silicon Layer*** (***ISL***) is a large radius ($R_{Min} = 22.6\ cm$, $R_{Max} = 29.0\ cm$) silicon tracker with a total active area of $\approx 3.5\ m^2$. It is composed of 296 basic units, called ladders, made of three silicon sensors bonded together in order to form one electric unit. Figure 1.c gives a schematic representation of the ISL detector. It is located between the Silicon Vertex Detector and the Central Outer Chamber. Being at a distance of $22.6 \div 29.0\ cm$ in the central part, from the beam-line, it covers a pseudo-rapidity region up to $|\eta| < 2$.

2. **Central Outer Tracker** (**COT**): that is the new CDF II central tracking chamber. It is an open cell drift chamber able to operate at a beam crossing time of $\approx 132\ ns$ with a maximum drift time of $\approx 100\ ns$. The COT consists of 96 layers arranged in four axial and four stereo superlayers. It also provides $dE/dx$ information for particle identification.

3. **Time-of-Flight Detector** (**TOF**): New scintillator based Time-Of-Flight detector has been added using the small space available between the COT and the solenoid. With *110 ps* time-of-flight resolution, the TOF system enhances the capability to tag charged kaons in the $P_T$ range from $\approx 0.6$ to few $GeV/c$ as requested from the B physics searches underway. Figure 3.a and 3.b show the TOF particle ID resolution as function of particle momentum. Figure 3.c and 3.d demonstrate how significantly change signal and background separation by adding TOF information in the $\Phi(1020)$ reconstruction.

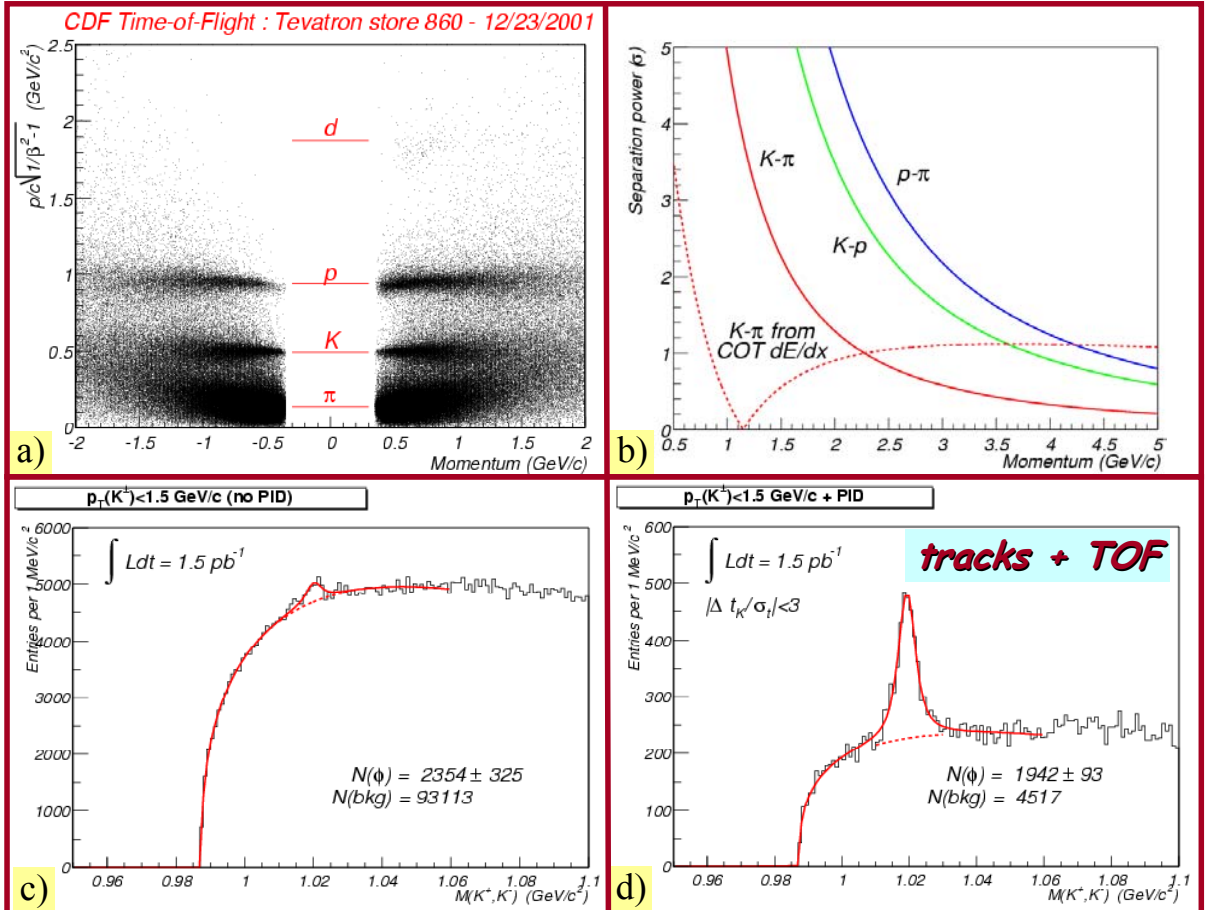

**FIG 3. a)** *and* **b)** *Expected TOF particle ID resolution as function of the particle momentum for low $P_T$ particles;* **c)** *$\Phi(1020)$ signal reconstructed with low momentum ($P_T < 1.5\ GeV/c$) kaon;* **d)** *the same as in c) after the TOF information have been used.*

4. **Plug Calorimeter**: A new scintillating tile plug calorimeter has been realized in order to have a good electron identification up to a pseudorapidity range of $|\eta| < 2$.
5. **Muon system** has also been upgraded. The coverage in the central region has been almost doubled compared to Run I situation. And a new forward detector, the (IMU) have been added.
6. **Trigger** The CDF II trigger is organized in 3 different levels. The Level 1 trigger (L1) is a dead-timeless trigger with a 42 stage pipeline and can make a trigger decision every 132 ns with a total latency time of 5544 ns. A new online processor reconstruct COT tracks (e**X**tremely **F**ast **T**racker). L2 trigger adds information within $\approx 20\ \mu s$, to the objects found by L1 trigger (electromagnetic or hadronic parts of the calorimeters, missing transverse energy, stubs in the muon system).
7. **Data Acquisition System (DAQ)**} has been adapted to short bunch spacing of 132 ns. It is capable to record data with event size of the order of 250 KB and permanent logging of 20 MB/s.

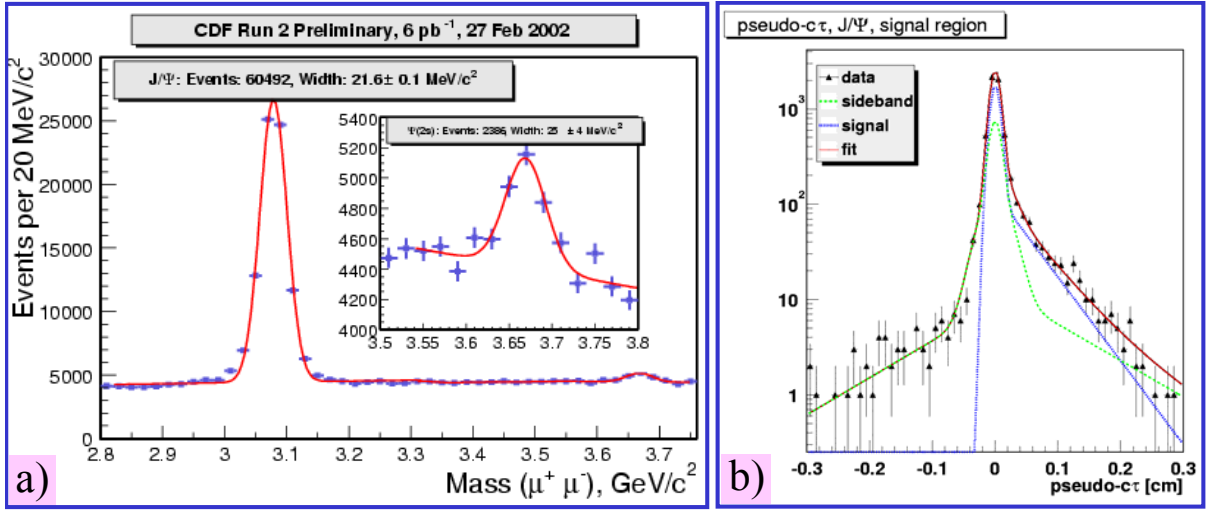

**FIG 4. a)** *Run II reconstructed $J/\psi$ and $\Psi(2s)$ events selected by using the dimuon trigger;* **b)** *$c\tau$ distance for the $J/\Psi$ sample.*

## 3. RUN II Early Physics Results

### 3.1 B Physics

B physics is an extraordinary laboratory to test several fundamental aspects of the Standard Model (SM). During the Run I, 1992-1996, 110 pb$^{-1}$ of data had been collected and used to perform important B physics measurements [3] including the first $\sin(2\beta)$ measurement on unitary triangle. The Tevatron collider, as mater of fact, is an excellent place for B physics studies both because it is possible to produce the full spectrum of mesons and baryons with b quarks, and also because the *b* hadrons production cross section is large (compare *100 mb* with few *nb* of $e^+e^-$ colliders). Many of the described detector upgrades have been done indeed with a large emphasis on heavy quark physics (*c, b, t*). CDF Run II physics program can be synthesized as follow. CP violation measurements using modes such as $B \to J/\psi K_s$, $B \to \pi\pi$, $B_s \to KK$ and $B_s \to D_s K$, $B_s$ Mixing, searches for rare B decays, measurement of lifetimes, masses and branching ratios. At present, the B physics program is going through the refining of triggers strategies and performing various high rate measurements. Later on, we plan to measure the $\sin(2\beta)$ along with a measurement of $B_s^0$ flavour oscillations by fully reconstructing $B_s^0$ decays ($B_s^0 \to D_s^-\pi^+$ and $B_s^0 \to D_s^-\pi^+\pi^-\pi^+$ with $D_s^-$ reconstructed as $\phi\pi^-$, $K^{*0}K^-$, $K_s^0 K^-$) and then, after more than $200 \div 300 \ pb^{-1}$ will be available, the study of rare decays, together with the refining of the previously performed measurements, will take over. Preliminary CDF Run II results are shown in Fig. 5 and in Table I.

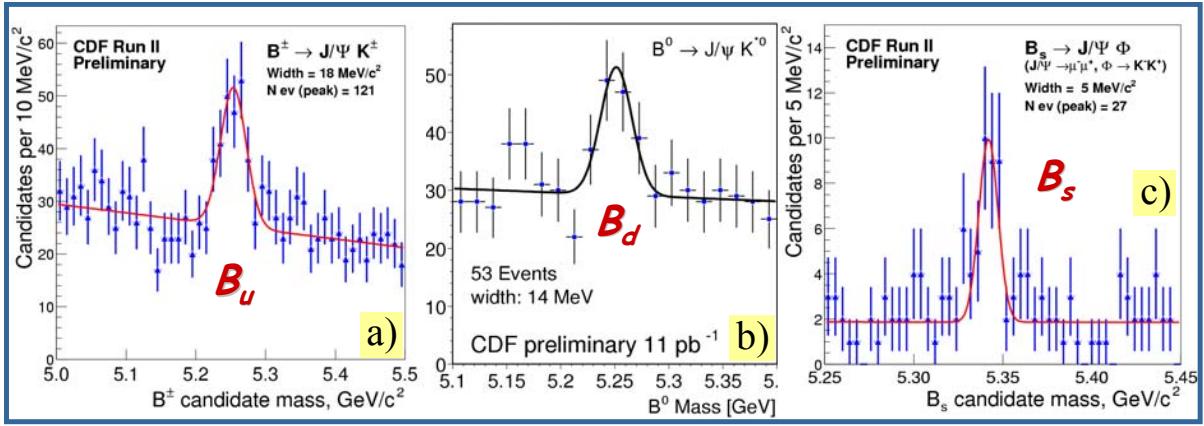

**FIG. 5.** *Candidate invariant mass for* **a)** $B_u^\pm \to J/\psi K^\pm$; **b)** $B_0^\pm \to J/\psi K^0$; **c)** $B_s^\pm \to J/\psi \Phi$.

Trigger Strategies

The total cross section for light quark production is 3 order of magnitude larger than b-quark production. B hadrons are then selected by using three general trigger strategies:
1. Hadronic Trigger;
2. Lepton plus displaced track Trigger;
3. Di-lepton Trigger:

The first trigger strategy take advantage of the long B hadrons lifetime to discriminate fully hadronic B decays from background. For the Run II CDF have been equipped with a Secondary Vertex Trigger (SVT) that selects events that pass at Level 1 the loose request of having two tracks with $P_T^{trk} > 2\ GeV/c$ in the event and, at Level 2, a displaced vertex, searched by requiring a large impact parameter: $D_0 > 100\mu m$. This trigger is extremely useful as it is able to selects both rare two body decays such as: $B \to \pi\pi$ (KK), relevant for CP violation measurement, as well as hadronic $B^\pm$ decays. The lepton plus displaced track trigger is a trigger that selects events containing electron or muons with $P_T^{trk} > 4\ GeV/c$ with the further request of an additional track with large impact parameter. Di-lepton Trigger looks for the presence of two opposite sign muons with $P_T^\mu > 1.5\ GeV/c$ or two opposite sign electrons with $P_T^e > 2.0\ GeV/c$. This trigger is relevant in selecting events as $B \to J/\psi K_S$ that will be used for measuring both $\sin 2\beta$ and the exclusive B meson lifetimes and also to search for EWK penguin decays such as $B \to K^{(*)}\ell^+\ell^-$ and other rare decays such as: $B_{(s)} \to \ell^+\ell^-$.

| B Meson | Mass(MeV/c²) | ΔPDG/σ(CDF) | σ(CDF)/σ(PDG) |
|---|---|---|---|
| $B_u^+ \to J/\psi K^+$ | 5280.6 ± 1.7 ± 1.1 | +0.77 | 4.05 |
| $B_d^0 \to J/\psi K^{0*}$ | 5279.8 ± 1.9 ± 1.4 | +0.17 | 4.72 |
| $B_s^0 \to J/\psi \varphi$ | 5360.3 ± 3.8 ± $^{2.1}_{2.9}$ | −1.81 | 1.90 |

**TABLE I.** *Comparison of the measured Run II meson masses with PDG values.*

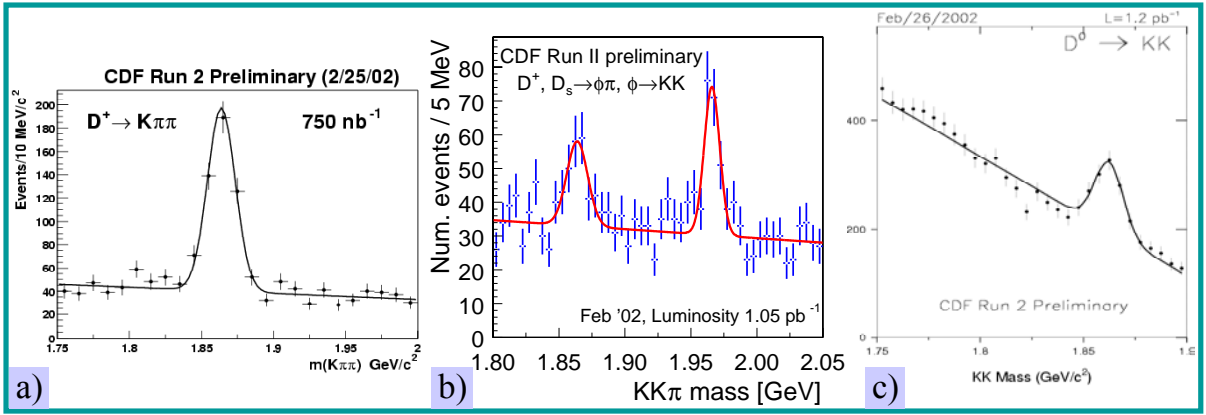

**FIG 6. a)** $D^*$ *signal reconstructed in the* $D^* \to K\pi\pi$ *decay mode;* **b)** $D^+$ *and* $D_s$ *signals reconstructed in the* $D^+, D_s \to \phi\pi, \phi \to KK$ *mode;* **c)** $D^0$ *signals reconstructed in the KK decay channel.*

## 3.2 Charm Physics

The SVT B trigger turned out to be extremely efficient also in selecting events enriched in charmed mesons. With the expected Run IIA integrated luminosity of $\approx 2\ fb^{-1}$ CDF II will be able to collect a charm sample up $10 \div 100$ time larger than those coming from fixed target experiments (see Table II). Large amounts of Cabibbo suppressed $D^0 \to K^+K^-$ decays and $D^0 \to \pi^+\pi^-$ are also observed. Other D species are also observed as shown in Fig. 6. Direct charm production is separated from charm coming for B meson decays by looking at the impact parameter. CDF II is at present measuring the differential production cross section for: $D^0$, $D^+$, $D^{*+}$, $D_s$ and $\Lambda_c$ hadrons. Charged modes such as $D^+ \to K^+K^-\pi^+$ and neutral modes such as $D^0 \to \pi^+\pi^-$ will be both used in order to search for direct CP violation.

| Decay Channel | | | Events |
|---|---|---|---|
| $D^0$ | $\to$ | $K\pi$ | $6 \times 10^5$ |
| $D^0$ | $\to$ | $KK$ | $4 \times 10^4$ |
| $D^0$ | $\to$ | $\pi\pi$ | $2 \times 10^4$ |
| $D^+$ | $\to$ | $K\pi\pi$ | $4 \times 10^5$ |
| $D_s$ | $\to$ | $KK\pi$ | $2 \times 10^4$ |
| $D^{*+}$ | $\to$ | $D^0\pi (D^0 \to K\pi)$ | $1.6 \times 10^4$ |
| $\Lambda_c$ | $\to$ | $pK\pi$ | 600 |

**TABLE II.** *Expected yield of hadronic charm decays with* $100\ pb^{-1}$ *of data.*

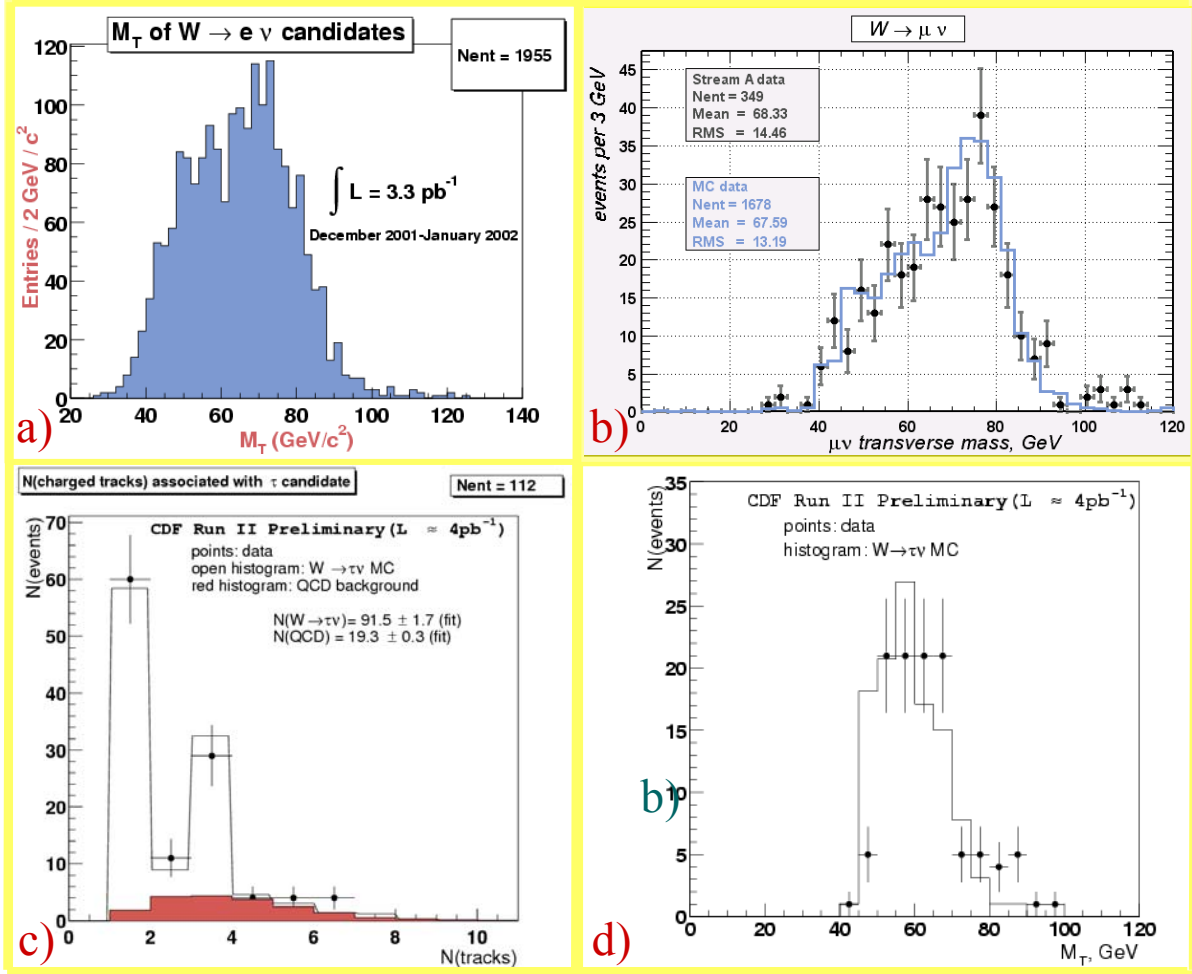

**FIG 7.** *Transverse mass distribution for:* **a)** $W \to e \nu_e$; **b)** $W \to \mu \nu_\mu$; **d)** $W \to \tau \nu_\tau$; **c)** *Number of tracks expected and observed inside the $\tau$ – jet for $W \to \tau \nu_\tau$ signal and for QCD background.*

### 3.3 Electroweak Physics

At Tevatron Collider, the W-bosons are produced by hard collisions between the constituent quarks and anti-quarks of the proton and anti-proton. During the Run I CDF measured the W boson mass with a precision of about $80 \, MeV/c^2$ leading to a combined CDF and DØ results of $M_W = 80.447 \pm 0.042 \, GeV/c^2$ that is the world's most precise measurement [4]. CDF II is expecting to measure the W mass with a precision of $\delta M_W \approx 30 \, MeV/c^2$ on an integrated luminosity of $2 \, fb^{-1}$. We started to look to all 3 leptonic decay channels of the W: $W \to \ell \nu_\ell$ ($\ell = e, \mu, \tau$) using the data collected till now. The W boson mass is extracted from fitting, with appropriate invariant mass, the transverse mass distribution defined as: $M_T \equiv \sqrt{\left(E_T^\ell + E_T^\nu\right)^2 - \left(\vec{p}_T^{\,\ell} - \vec{p}_T^{\,\nu}\right)}$. Figure: 7.a, 7.b and 7.d show the W transverse mass distributions as obtained in the case of $W \to e\nu_e$, $W \to \mu\nu_\mu$ and $W \to \tau\nu_\tau$. In figure 7.c we compare the number of tracks expected in the $\tau - jet$ for $W \to \tau\nu_\tau$ signal and for the QCD background (MC) with the $W \to \tau\nu_\tau$ from the data. Events are selected by requiring an isolated electron or muon with $P_T > 25\, GeV/c^2$ and the presence of a consistent amount of missing transverse energy $\not{E}_T > 25\, GeV$. By analyzing $10\, pb^{-1}$ we found 5547 W. The W signal is quite clean indeed we expect the background fraction not to exceed $\approx 7\%$. This sample have been used also to estimate the production cross section: $\sigma_W \times BR(W \to e\nu_e) = 2.60 \pm 0.03\,(Stat) \pm 0.26\,(Lu\min osity)\, nb$ and found to be consistent with the Run I measurement.

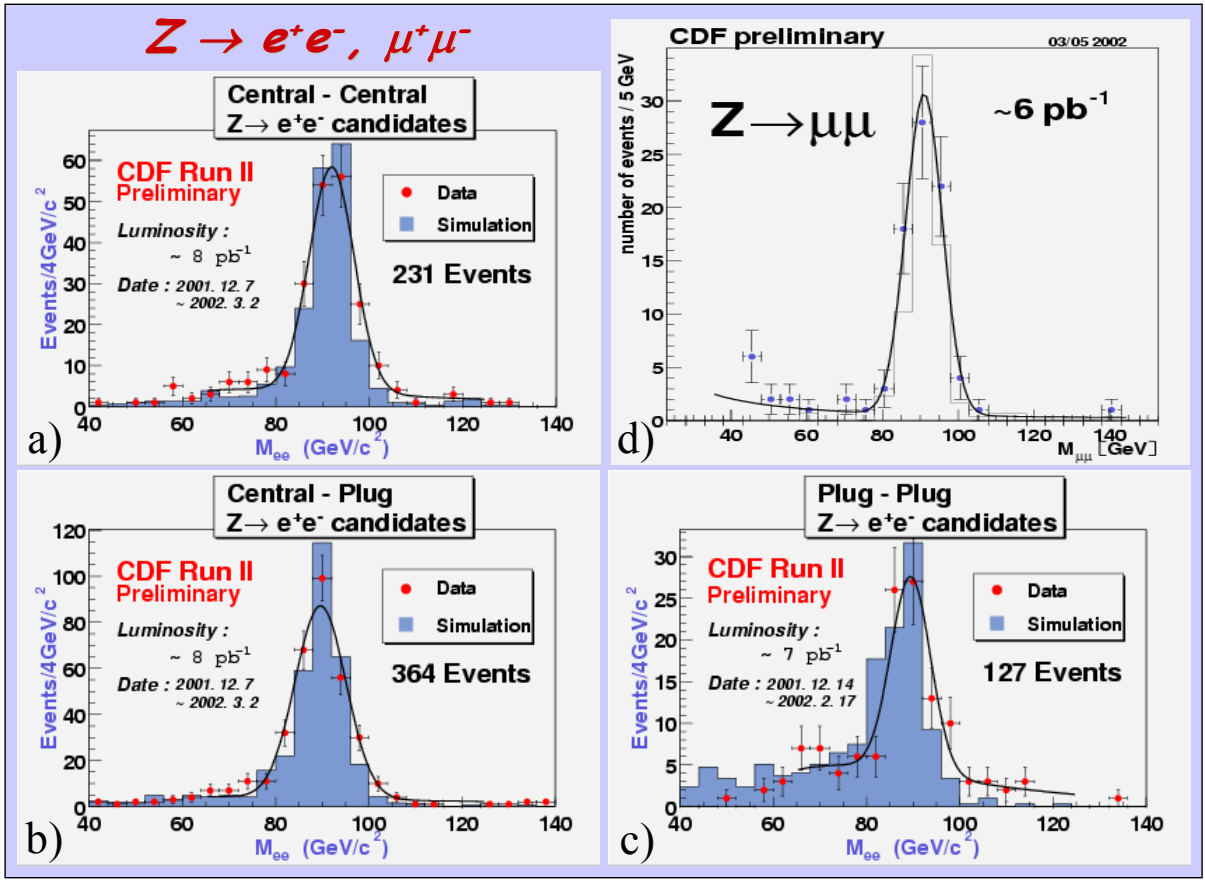

**FIG 8**. $Z \to \ell^+\ell^-$ invariant mass peak as observed in: **a)** central-central electron sample, **b)** central-plug electron sample, **c)** plug-plug electron sample and **d)** in the $\mu^+\mu^-$ channel.

We have already clean $Z \to e^+e^-$ and $Z \to \mu^+\mu^-$ samples. Figure 8.a, 8.b, 8.c show the $Z \to e^+e^-$ invariant mass as reconstructed for central-central electrons, central-plug electrons and using electrons having both the legs into the plug detector. Fig 8.d shows the $Z \to \mu^+\mu^-$ invariant mass as reconstructed on a data sample of $6\, pb^{-1}$. Analysis on the $Z \to \tau^+\tau^-$ channel is underway and, also for this analysis, preliminary results will viable for the winter conferences.

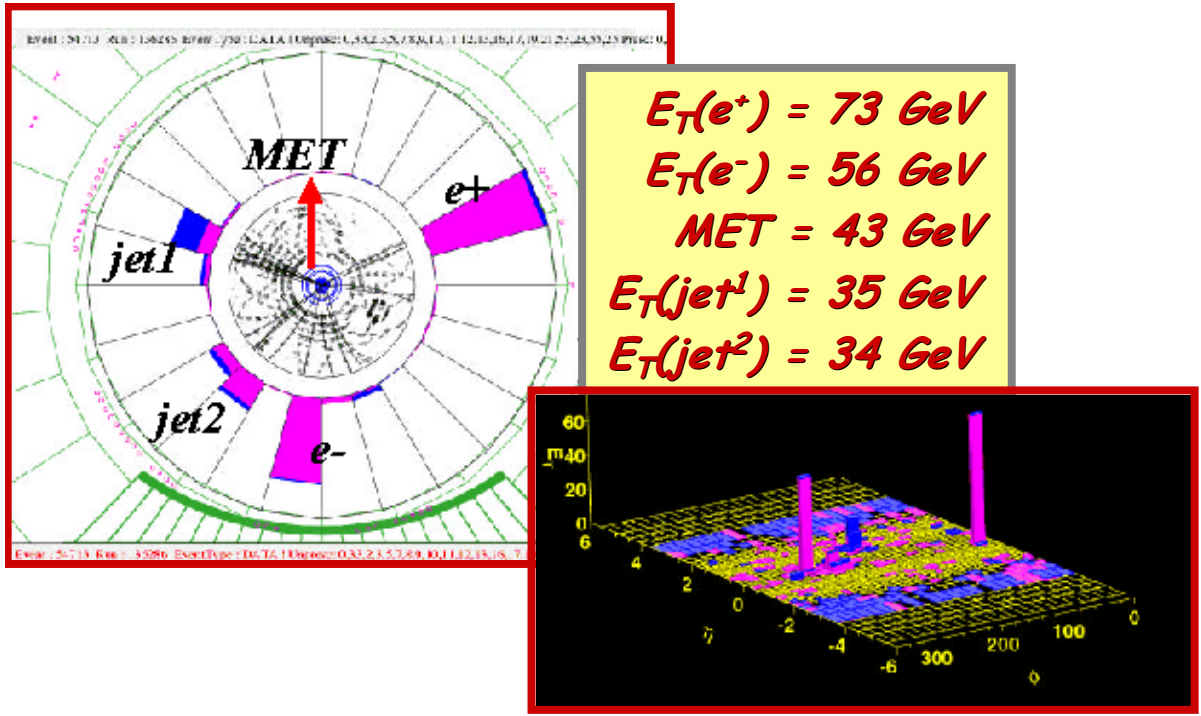

**FIG 9.** Run *II dielectron Top candidate as displayed using the CD II event display.*

3.4 Top Quark Physics

At Tevatron, top quarks are predominantly pair-produced, with each top quark decaying to a $W$ and to a $b$ quark: $t\bar{t} \rightarrow WbWb$. The increase in the center of mass energy from 1.8 *TeV* to 1.96 *TeV* was mainly motivated by the consistent rise of the top quark production cross section of $\approx 30 \div 35\%$. CDF II is expecting to collect a top quark sample $\approx 30$ times bigger than in run I, assuming an integrated luminosity of $\approx 2\,fb^{-1}$. At present we are finalizing important tools such as the b-tagging based both on jet probability and on b hadron vertex displacement and the jet corrections. CDF II can expect to measure, using the first $2\,fb^{-1}$ the top quark mass with a precision of $\delta M_t \approx 3\,GeV/c^2$. During the Run I the combined CDF and DØ top quark mass measurement was: $M_t = 174.3 \pm 5.1\,GeV/c^2$ [5]. Combined knowledge of $M_t$ and $M_W$ will allow us to set a more stringent limit on the Standard Model Higgs mass [6]. In figure 9 we show one of the Run II di-electron top candidates ($t\bar{t} \rightarrow W^+W^-bb \rightarrow e^+\nu_e e^-\bar{\nu}_e bb$). Both in the central cutaway and lego plot is possible to see the two candidate electron and jets present in the final state.

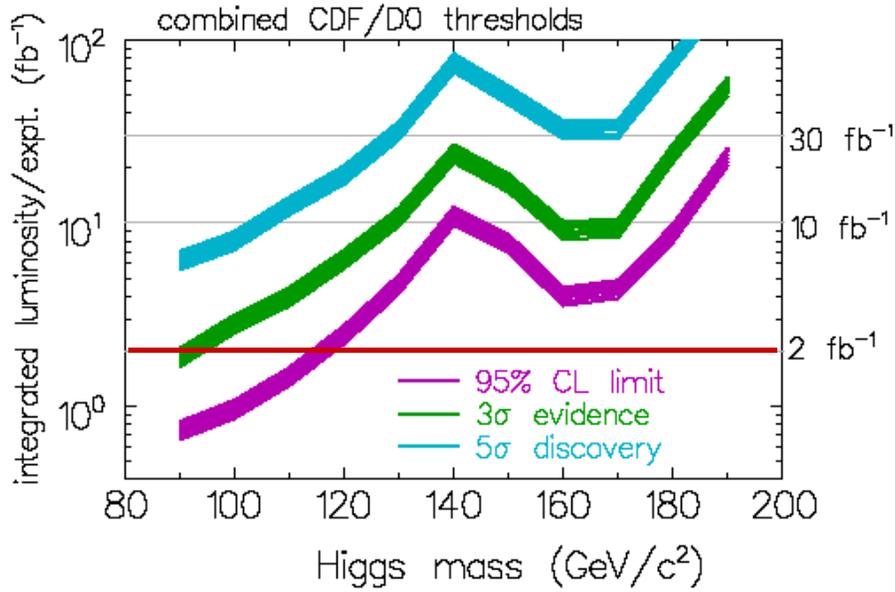

**FIG 10.** *Integrated luminosity required per experiment to either exclude at 95% C.L. or discover with a $3\sigma$ or $5\sigma$ significance a SM Higgs boson.*

## 4. Higgs Potential

At Tevatron the Higgs boson is expected to be produced mainly via gluon fusion or in association with $W$ or $Z$ bosons. Although the gluon fusion mode gives the most important contribution to the Higgs production, it will be overwhelmed by the large QCD background. Therefore, given sufficient luminosity, the most promising SM Higgs discovery mechanism for $m_H < 130 \, GeV/c^2$ consists of $q\bar{q}$ annihilation into a virtual $V^*$ ($V = W, Z$), where the virtual $V^* \to Vh_{SM}$ followed by $h_{SM} \to b\bar{b}$ and the leptonic decay of the V that will serve as a trigger. The main background for this mode will be $Wb\bar{b}$ and $WZ$ processes. For $120 \, GeV/c^2 < m_H < 190 \, GeV/c^2$, where the Higgs is produced with a vector bosons, it will mainly decay into $W^*W^*$ states with subsequent decay $(W,Z)W^*W^* \to \ell^\pm \nu \ell^\pm \nu jj$. For this case selection criteria requires two leptons with $P_T > 10 \, GeV/c$ having the same charge and two separate jets with $E_T^{jet} > 15 \, GeV/c$ and the presence of missing transverse energy. The main background in this case is $WZjj$ production. Among various analyses underway some interesting result could also come from the use of neural networks techniques. The integrated luminosity required per each Tevatron experiment, to exclude a 115 $GeV/c^2$ SM Higgs boson at 95% C.L. is 2 $fb^{-1}$.

## 5. Conclusions

In this paper we reviewed recent CDF II results reporting on the status of the detector and its upgrades. The CDF II detector is presently performing very well and has collected until the middle of June 50 pb$^{-1}$ of data. We expect by the end of this year to reach an integrated luminosity of $\approx 200 \ pb^{-1}$. The understanding of the detector performances is very advanced so many physics analysis are in progress. With the new detector capabilities a broad physics program is within our reach. We expect to present new interesting results by the winter 2003.

## 6. Acknowledgments

I wish to thank the Organizers of the SUSY02 Conference for the excellent conference and their kind hospitality.